\documentclass[useAMS,usegraphicx,usenatbib]{mn2e}

\def\ltsima{$\; \buildrel < \over \sim \;$}
\def\simlt{\lower.5ex\hbox{\ltsima}}
\def\gtsima{$\; \buildrel > \over \sim \;$}
\def\simgt{\lower.5ex\hbox{\gtsima}}
\def\kpc{{\rm\,kpc}}

\makeatletter
\newcommand\ion[2]{#1$\;${\small\rmfamily\@Roman{#2}}\relax}%
\makeatother

\def\s{\ifmmode \widetilde \else \~\fi}
\def\={\overline}

\def\spose#1{\hbox to 0pt{#1\hss}}

\def\lta{\mathrel{\spose{\lower 3pt\hbox{$\mathchar"218$}}
     \raise 2.0pt\hbox{$\mathchar"13C$}}}
\def\gta{\mathrel{\spose{\lower 3pt\hbox{$\mathchar"218$}}
     \raise 2.0pt\hbox{$\mathchar"13E$}}}
\def\Dt{\spose{\raise 1.5ex\hbox{\hskip3pt$\mathchar"201$}}}    
\def\dt{\spose{\raise 1.0ex\hbox{\hskip2pt$\mathchar"201$}}}    

\def\dotsfill{\leaders\hbox to 1em{\hss.\hss}\hfill}

\title{A panoramic view of M81: New stellar systems in the debris field}

\author[Mouhcine \& Ibata]{M.~Mouhcine$^1$, R.~Ibata$^2$ \\
$^1$Astrophysics Research Institute, Liverpool John Moores University, 
    Twelve Quays House, Egerton Wharf, Birkenhead, CH41 1LD, UK \\
$^2$Observatoire Astronomique de Strasbourg (UMR 7550),
    11, rue de l'Universit\'e, 67000 Strasbourg, France }

\date{Accepted ?. Received ?; in original form ?}

\pagerange{\pageref{firstpage}--\pageref{lastpage}}
\pubyear{2007}

\begin{document}

\maketitle

\label{firstpage}

\begin{abstract}

Using the MegaCam imager on the Canada-France-Hawaii Telescope, we have 
resolved individual stars in the outskirts of the nearby large spiral galaxy 
M81 (NGC~3031) well below the tip of the red giant branch of metal-poor 
stellar populations over $\sim 60 \kpc \times 58 \kpc$. In this paper, we 
report the discovery of new young stellar systems in the outskirts of M81. 
The most prominent feature is a chain of clumps of young stars distributed 
along the extended southern H{\sc i} tidal arm connecting M~81 and NGC~3077. 
The colour-magnitude diagrams of these stellar systems show plumes of bright 
main sequence stars and red supergiant stars, indicating extended events 
of star formation. The main sequence turn-offs of the youngest stars in the 
systems are consistent with ages of $\sim 40$ Myr. The newly reported stellar 
systems show strong similarities with other known young stellar systems in 
the debris field around M81, with their properties best explained by these 
systems being of tidal origin.

\end{abstract}

\begin{keywords}
galaxies: formation -- galaxies: stellar content -- 
galaxies: individual (M81, NGC~3077, Holmberg IX, BK 3N, Garland) -- 
galaxies: photometry
\end{keywords}

\section{Introduction}
\label{intro}

\footnotetext[1]{This publication is based on observations with the 
MegaPrime/MegaCam, a joint project of the CFHT and CEA/DAPNIA, at the
Canada-France-HAwaii Telescope (CFHT), which is operated by the National
Research Council of Canada, the institut National des Sciences de 
l'Univers of the Centre National de la Recherche Scientifique (CNRS), 
and the University of Hawaii.}

In recent years, it has been increasingly recognised that the outskirts 
of galaxies hold fundamental clues about their formation history. 
It is into these regions that new material continues to arrive as part 
of their assembly, by accretion of minor satellites, predominantly at 
early epochs when large disk galaxies were assembling, as predicted by
the currently favored hierarchical formation models.
It was also in the outer regions of galaxies that material was deposited 
during the violent interations in the galaxy's past. Most present-day 
disk galaxies are suspected to have experienced mergers during the 
last few billions of years \citep[e.g.][and references therein]{hammer07}. 
Based on the systematic deviation of the Milky Way from a number of 
galaxy scaling relations, \citet{hammer07} have agrued that our galaxy 
had most likely escaped any significant major merger event over the 
last $\sim 10$ Gyr. These authors suspect that the observed differences 
between the Milky Way and its neighbour M31 are likely due to the 
quiescent formation in the former case and to the merger-dominated 
history for the latter. The observed properties of the stellar content 
in the outskirts of M31 can be accounted for by either a succession of 
minor mergers or a major merger, with this material most likely accreting 
in the most recent half of the age of the Universe \citep{ibata05}.

By analysing the characteristics of spiral galaxy stellar halos 
formed within a large grid of numerical chemo-dynamical simulations,
\citet{renda05} have shown that at any given total galactic mass, 
the metallicities of simulated stellar halos span a range in excess 
of $\sim$~1~dex. The underlying driver of this metallicity spread can 
be traced back to the diversity of galactic mass assembly histories. 
Galaxies with a more extended merging history possess halos which have 
younger and more metal rich stellar populations than the stellar halos 
associated with galaxies with a more abbreviated assembly. For a given 
total mass, galaxies with more extended assembly histories also possess 
more massive stellar halos.

The studies of the Galaxy, and to lesser extent the other large spiral in 
the Local Group, M31, have been delivering the bulk of the observational 
constraints on the properties of the stellar content of the outer regions 
of galaxies. Evidence indicates that the Galaxy might be unrepresentative 
of a typical spiral galaxy, and it may not even follow the standard 
scenario of disk formation. The Milky Way halo seems to be populated by 
old, metal-poor stars, while a few fields in the halo of M31 show a large 
population of intermediate age stars with a much higher overall metallicity 
\citep{brown06}. The fields in M31 in which the above results were obtained 
have been found to be significantly contaminated by various accretion 
events \citep{ibata07} casting doubt on the conclusion that the M31 halo 
is globally younger and more metal-rich than that of the Milky Way. 
The current observational evidence therefore demonstrates that halos are 
complex structures. To establish comprehensively properties of stars in 
the outskirts of galaxies, and to fully understand their nature and origin, 
we need to undertake panoramic studies of the outer regions of spiral 
galaxies beyond the Local Group. To do so, we have obtained deep and 
wide-field optical imaging data of the nearby early-type spiral M81 
(NGC~3031), resolving stars well below the tip of the red giant branch 
of metal-poor stellar populations. 

The M81 group of galaxies is one of the nearest groups to our own. 
It contains one large spiral, two peculiar galaxies (M82 and NGC~3077), 
two small spirals galaxies (NGC~2976 and IC~2574), as well as a large number 
of dwarf galaxies \citep[e.g.][]{karachentsev85, karachentsev01, chiboucas08}. 
The core galaxies of the group are strongly interacting. Atomic hydrogen 
observations have revealed the presence of a large number of tidal streams 
with large, dynamically complex atomic hydrogen clouds embedding M81, 
M82, NGC~3077, and NGC~2976 \citep{vdh79,appleton81,yun94,boyce01}. 
Close interactions between galaxies are capable of leaving tidal debris 
that could be converted into new stellar systems \citep[e.g.][]{TT72}. 
Compared to the Local Group, an interesting feature of the M81 group is 
the presence of a population of stellar systems dominated by young stars 
\citep[e.g.][]{durrell04,demello08a,davidge08}, which are suspected to be of 
tidal origin \citep[e.g][]{makarova02}, and which have no counterparts in the 
Local Group. 
These young stellar systems, e.g. Holmberg IX, BK 3N, and Garland, are 
embedded in H{\sc i} clouds \citep{boyce01}. Here, we take advantage of 
our deep and wide field survey to study the spatial distribution of young 
stellar populations in the outer regions of M81. We report the discovery 
of new stellar systems in the tidal debris.

Analysis of the spatial distribution of old stellar populations, the 
bi-dimensional distribution, the search for substructures, the metallicity 
distribution functions, and globular cluster properties over the surveyed 
area will be reported in forthcoming papers. The stellar populations of 
immediate interest to the present paper are revealed by the upper main 
sequence and the red supergiant stars. The layout of this paper is as 
follows: in Section \ref{data} briefly represents the data set, while 
section \ref{results} studies the young stellar content around M~81.

\section{Data}
\label{data}

\begin{figure}
\includegraphics[clip=,width=0.5\textwidth]{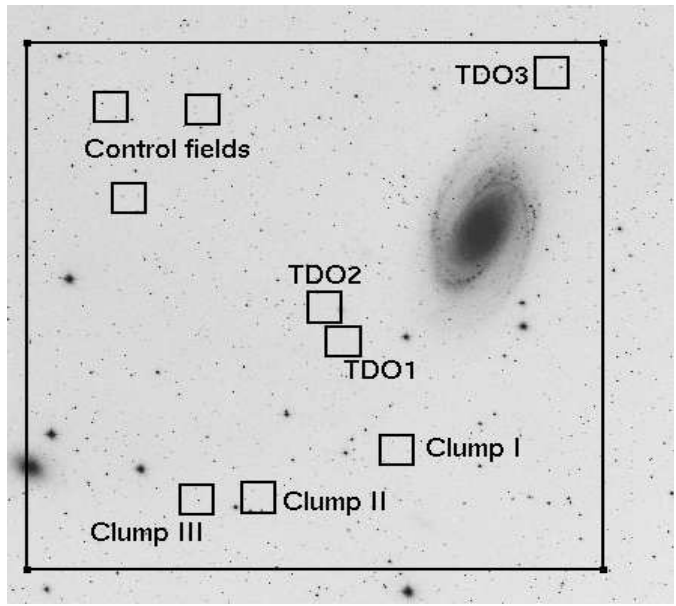}
\caption{CFHT/MegaCam footprint of the observations overlayed on the 
Digitized Sky Survey image of M~81. The MegaCam field-of-view covers a 
$0.96^{\circ} \times 0.94^{\circ}$ field. The three new young stellar clumps 
reported here are identified as Clump I, II, III (see text for more details). 
Also shown are the three tidal debris objects (TDO1, TDO2, TDO3) identified 
by \citet{davidge08}. The locations of three control fields are also shown.}
\label{footprint}
\end{figure}

The MegaCam wide-field imager at the Canada-France-Hawaii Telescope (CFHT) 
was used to map M81 over the area displayed in Figure \ref{footprint}. 
MegaCam consists of a mosaic of thirty six $2048\times 4612$ EEV chips, 
covering a $0.96^{\circ} \times 0.94^{\circ}$ field, with each pixel subtending 
0.187 arcseconds on a side. The photometric depth and field of view achievable 
with this instrument make it particularly powerful in regions of extremely low 
stellar surface density. To observe as much of the galaxy halo and the extension 
of the galaxy disk as possible, we positioned the centre of M81 in one corner 
of the mosaic. The images used in this study are of a single pointing centred 
at ${\rm R.A.} = 09^{\rm h}58^{\rm m}44.0^{\rm s}$, 
${\rm Dec.} = +68\degr 51\arcmin 46.0\arcsec$ (J2000). The survey gives an 
uninterrupted coverage out to approximately 50 kpc along the major axis, and 
the inner halo out to $\sim 45$ kpc. 

The observations were taken with the $g$ and $i$ filters, with total exposure 
times of 14\,000 and 20\,000 seconds, respectively, in each of these two bands, 
to reach $g\simeq 27.3$ and $i\simeq 25.9$. At the distance of M81, taken to 
be 3.55 Mpc \citep{freedman94} throughout this paper, we detect approximately 
the top 1.5 magnitudes of the red giant branch of metal-poor stellar 
populations. The data were obtained in dark skies, with typical seeing of 
0.9 and 0.8 arcseconds in the g- and i-bands, respectively (the relatively 
poor seeing is due to the low elevation of the target as observed from Hawaii). 
The individual exposures were recorded with a square-shaped dither pattern to 
assist with the identification of bad pixels and the suppression of cosmic 
rays. 

\begin{figure*}
\includegraphics[clip=,width=0.45\textwidth]{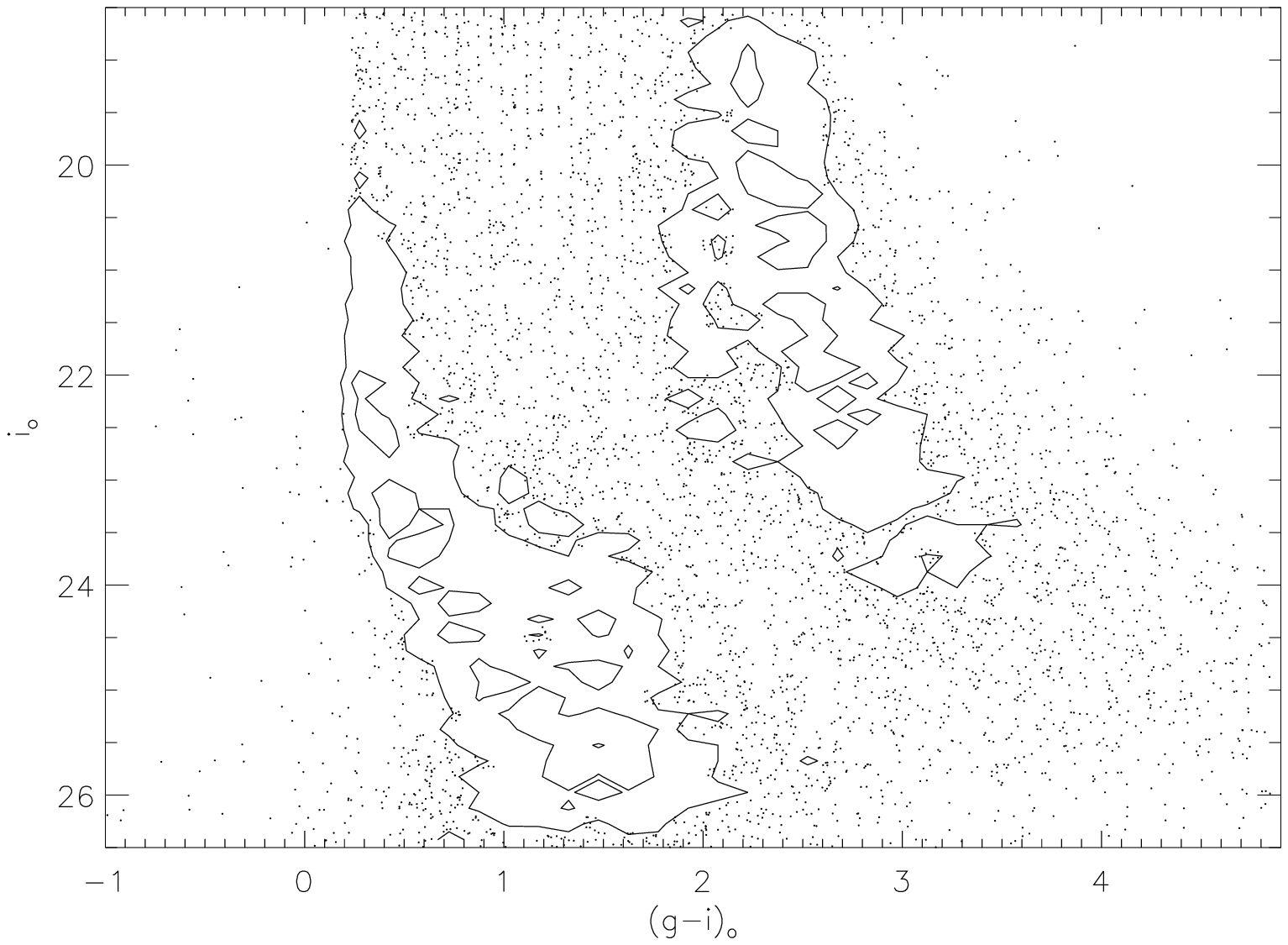}
\includegraphics[clip=,width=0.45\textwidth]{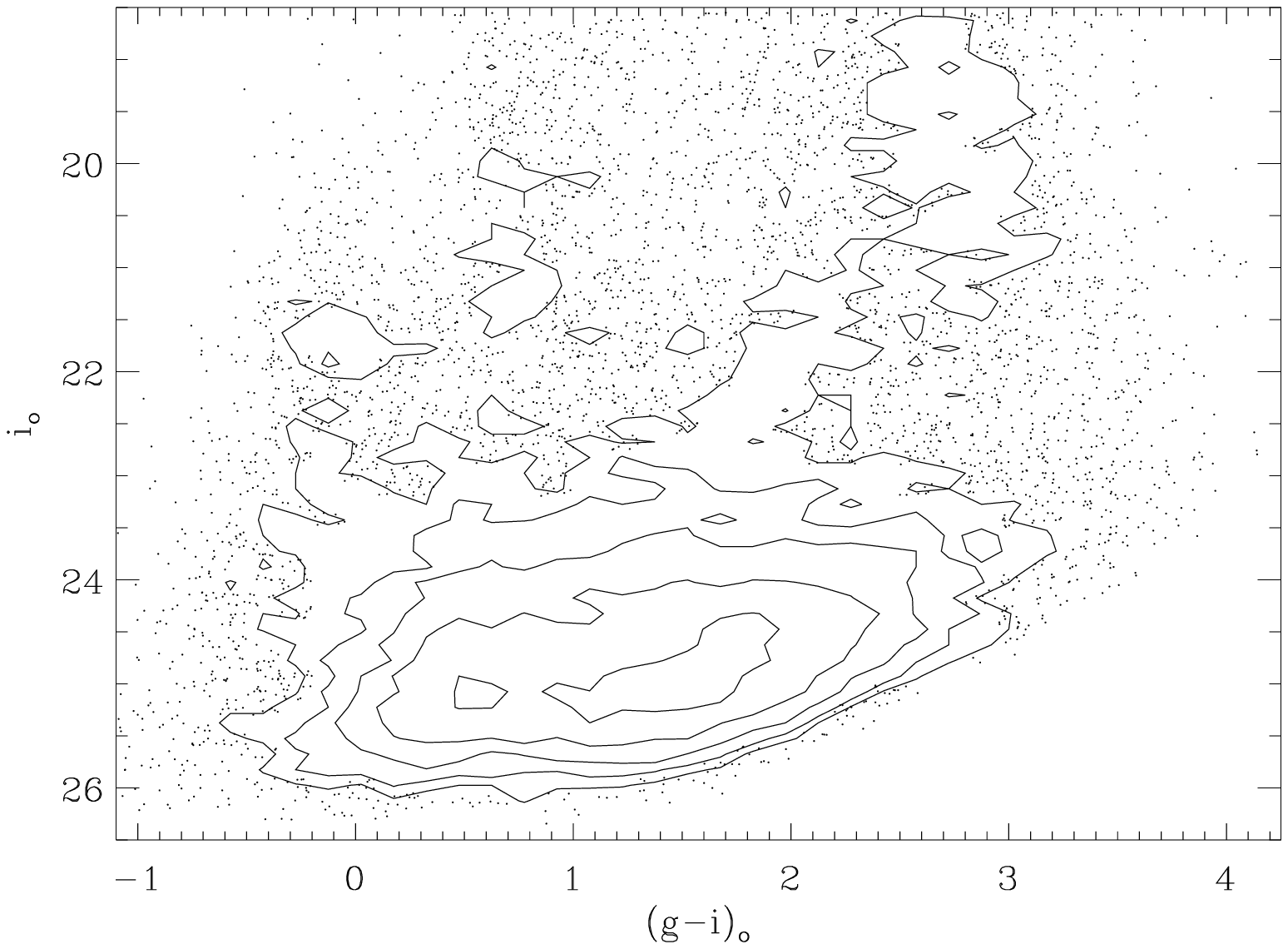}
\caption{Left: The $i_{\circ}$ versus $(g-i)_{\circ}$ CMD of foreground stars 
as predicted by The Besan\c{c}on Galactic population model in the direction 
of M~81 over the same field-of-view as covered with a single MegaCam pointing. 
Right: The $i_{\circ}$ versus $(g-i)_{\circ}$ CMD of objects detected in our 
field and classified as stars.  Regions of density less than ten stars per 
bin of $0.15 \times 0.15$ magnitudes are plotted as points. Contours are 
spaced by factors of two.}
\label{cmds}
\end{figure*}

The images were pre-processed by the CFHT Elixir pipeline for corrections 
for bias, flat-fielding, and the fringing pattern. Photometric standards 
observed over the season are used to determine the photometric zero point 
in each passband. The images were then processed by the Cambridge Astronomical 
Survey Unit (CASU) photometry pipeline \citep{irwin01}, in an identical manner 
to that described in \citet{segall07}. The interested reader is referred to 
this paper for more details. The software then proceeds to detect sources 
and measures their photometry, the image profile, and shape. Based on the 
information contained in the curve of growth, the algorithm classifies the 
objects into noise detections, galaxies, and probable stars. Throughout the 
paper we select as stars objects that have classification of -1 and -2 in both 
bands. This corresponds to stars up to $2\,\sigma$ from the stellar locus. 
To correct for the foreground extinction, we used the \citet{schlegel98} dust 
map value of $E_{(B=V)}=0.08$, corresponding to $A_g = 0.303$ and $A_i = 0.167$ 
respectively.

\section{Results}
\label{results}

\begin{figure}
\includegraphics[clip=,width=0.45\textwidth]{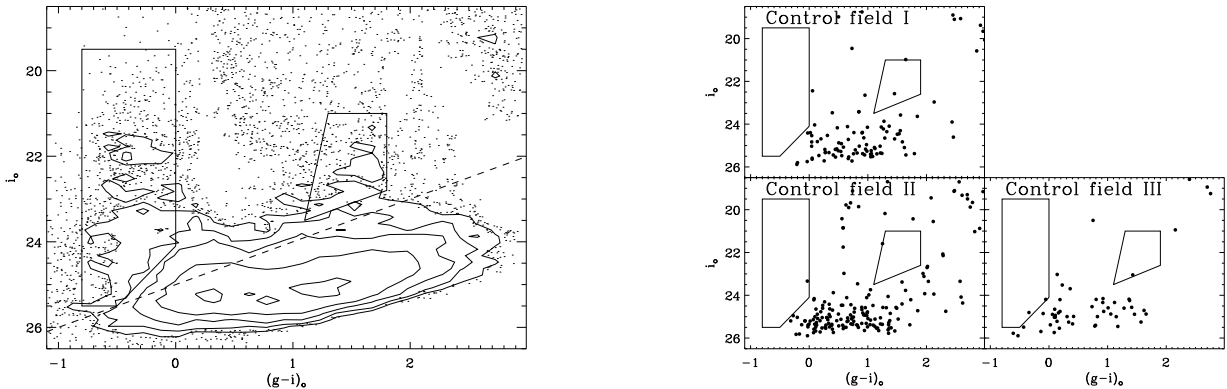}
\caption{Foreground-corrected CMD of stars in the MegaCam pointing. 
The solid boxes mark the locations of the colour-magnitude diagram where 
blue main sequence and red supergiant star candidates lie respectively.}
\label{cmd_fg_corrected}
\end{figure}

\begin{figure*}
\includegraphics[clip=,width=0.45\textwidth]{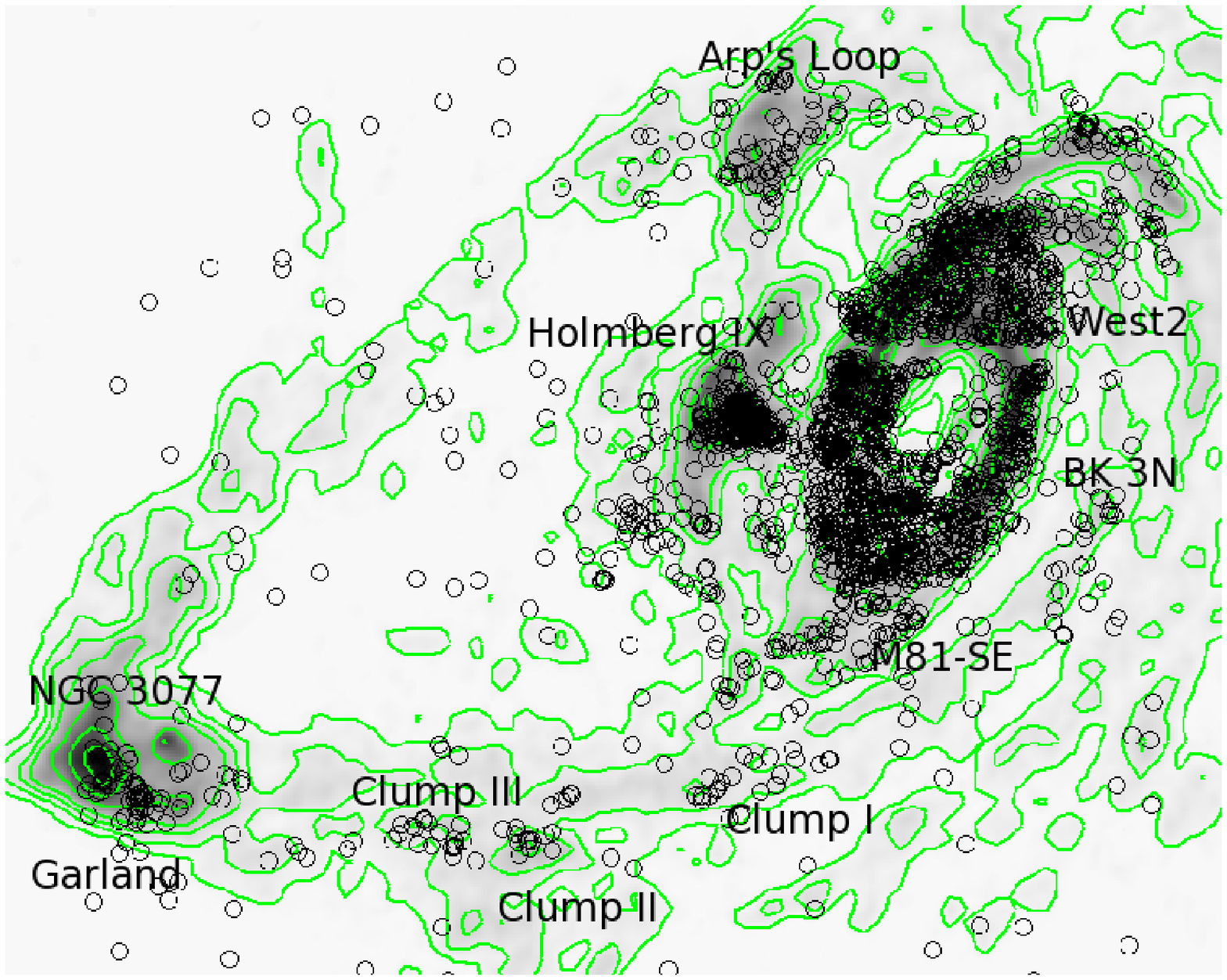}
\includegraphics[clip=,width=0.45\textwidth]{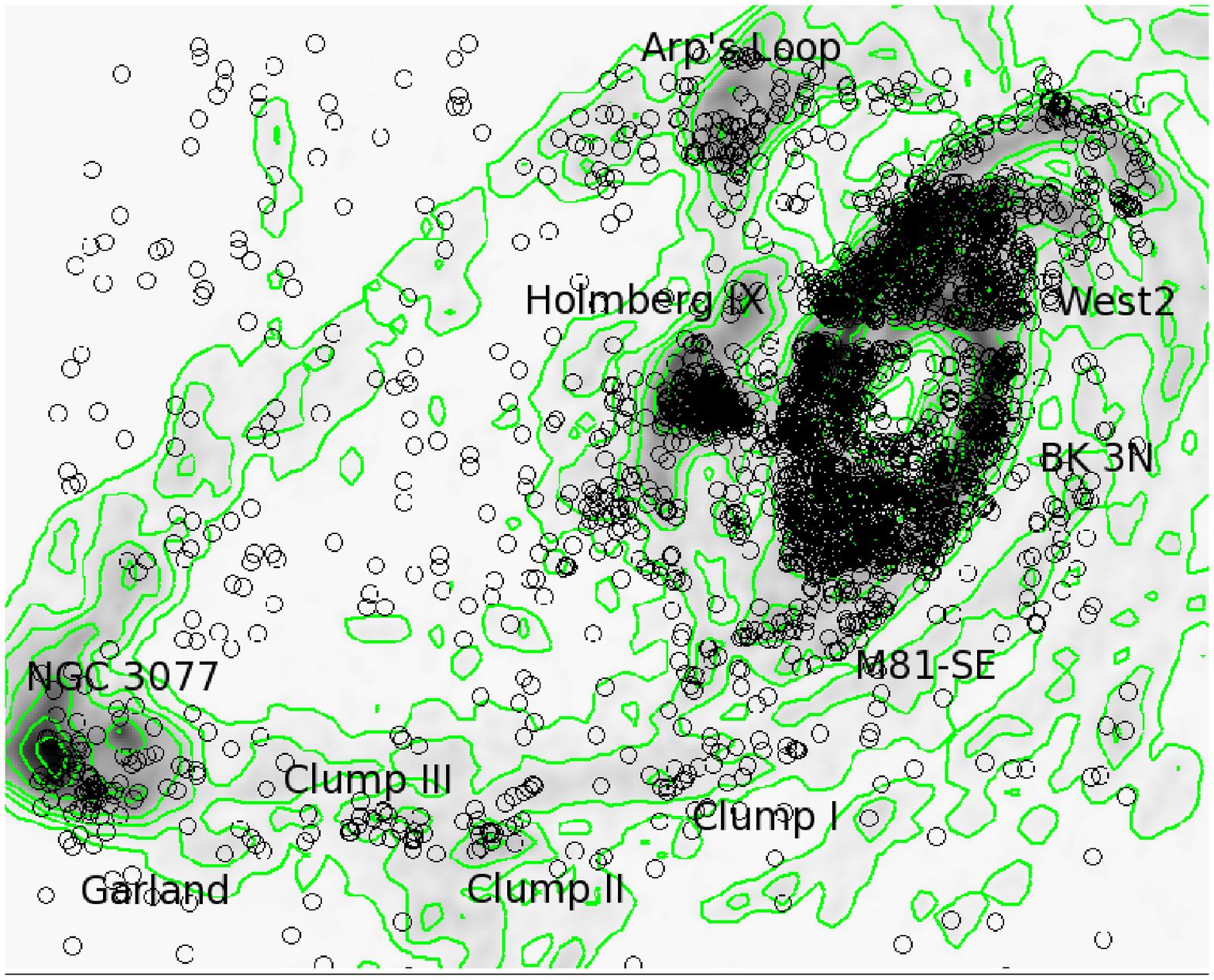}
\caption{(Left) Spatial distribution of main sequence star candidates, 
shown as open circles, over the MegaCam pointing. The contours and 
grayscale image show the H{\sc i} radio map from Yun et al. (1994). 
(Right) similar to the left panel, but for both main sequence and red 
supergiant star candidates. The H{\sc i} image adopted from Yun et al.'s 
Figure 1 by permission from Macmillan Publishers Ltd: Nature, copyright 
(1994). }
\label{young_stars_map}
\end{figure*}

\begin{figure}
\includegraphics[clip=,width=0.45\textwidth]{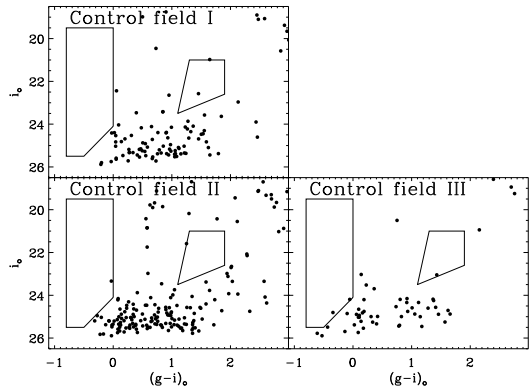}
\caption{Foreground-corrected CMD of stars in the three control fields. 
The solid boxes mark the locations of the colour-magnitude diagram where 
blue main sequence and red supergiant star candidates lie respectively.}
\label{cmd_control_field}
\end{figure}

\begin{figure}
\includegraphics[clip=,width=0.5\textwidth]{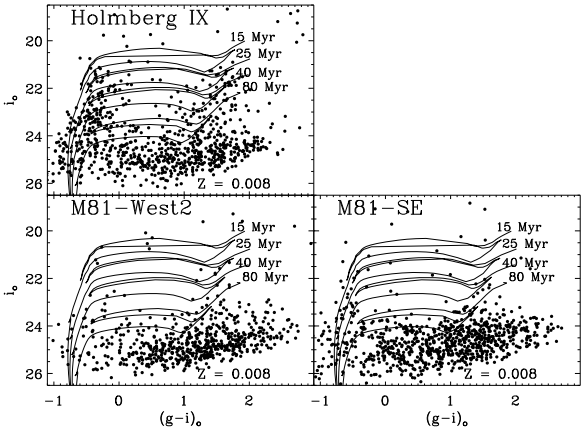}
\caption{The CMDs of stellar populations tracing the continuations of 
both the nourthen and the southern H{\sc i} spiral arms as identified 
in Fig.\,\ref{young_stars_map}. Superimposed on each CMD are the Padova 
isochrones with a metallicity Z=0.008 and the indicated ages. 
For a comparison purpose, the CMD of Holmberg IX is shown in the upper 
left panel. }
\label{cmd_sp}
\end{figure}

As well as encompassing a large fraction of M81 and its outskirts, the 
survey also intersects a non-negligible volume of the foreground Galaxy. 
To subtract off the foreground counts, we used the Besan\c{c}on Galactic 
population model \citep{robin03} to predict the foreground contamination. 
This Galaxy model has shown its capacity to predict reliably the observed 
foreground counts over large areas (see, e.g. \citealt{ibata07}). 
To substract the background counts, we tessellated the survey area with 
$0.25^{\circ}\times0.25^{\circ}$ bins and generated simulated catalogues 
using the Besan\c{c}on model. 
All stellar populations in the models with $i$-band magnitudes between 
$17<i_{\circ}<26$ were accepted. To reduce noise in the randomly generated 
catalogues, at each spatial bin we simulated a 30 times larger solid angle 
and later corrected the density maps for this factor. Finally, the artificial 
photometry was convolved with the observed magnitude-dependent error function. 
The left panel of Fig.~\ref{cmds} shows the predicted 
$i_{\circ}$ vs. $(g-i)_{\circ}$ colour-magnitude diagram (hereafter CMD) of 
the Galaxy foreground stars in the direction of M81. Regions of the CMDs 
with stellar densities higher than ten stars in a $0.15 \times 0.15$ 
magnitude bin are shown as contours with the contour levels spaced uniformly 
by a factor of two. Stars are plotted as points when the densities are lower. 
The foreground contamination is dominated by a prominent blue sequence, 
at $0 \la (g-i)_{\circ} \la 0.8$, of halo stars at or close to the main 
sequence turn-off at increasing distance through the Galactic halo. 
On the red side of the CMD, a prominent nearly vertical sequence at 
$2\la {g-i}_{\circ}\la 3$ and $19\la i_{\circ}\la 24$ is visible; the sequence 
is the result of Galactic disk dwarf stars accumulating over a large 
range of distances along the line of sight. 

The right panel of figure \ref{cmds} shows the combined CMD of all objects 
detected in our images and classified as stars in the deep MegaCam field. 
The CMDs are shown as density contours to reveal features in otherwise 
crowded regions. In addition to the blue and red sequences of foreground 
contaminants, a prominent vertical sequence of stars bluer than stars 
populating the blue foreground sequence, i.e., $(g-i)_{\circ} \le 0$, and 
covering a wide range of magnitudes, i.e., $19 \la g_{\circ} \la 26$, is 
present. A second sequence of bright, i.e., $21 \la i_{\circ} \la 22$, and 
red, i.e., $1.5 \la (g-i)_{\circ} \la 2.5$, stars is present. A third CMD 
feature is revealed by red giant branch stars.

Figure \ref{cmd_fg_corrected} shows the foreground-subtracted CMD of the 
stellar population distributed over the MegaCam field. The CMD is typical 
of stellar systems with extended star formation histories, with a prominent 
young stellar component. We see the upper main sequence and probable 
helium-burning blue stars, the red supergiant branch, and probably some 
young and intermediate age asymptotic giant stars. At $i_{\circ}\sim 25$, 
the contamination is dominated by background compact galaxies, unresolved 
in ground-based images and prone to misclassification as stellar objects. 
Characterising and correcting for the background contamination at faint 
magnitudes is beyond the scope of the present contribution, and will be 
addressed in a forthcoming paper. To reduce the contamination from 
background objects, we have selected stars brighter than $g_{\circ}=25$. 
A second potential source of contamination when tracing the spatial 
distribution of young stars in the debris field is the population of 
intermediate and old stars, either genuine members of the tidal debris 
or belonging to M81 and/or NGC~3077. These stars start to dominate at 
$i_{\circ}\sim 24$ and $(g-i)_{\circ}\sim 1$. We have excluded stars fainter 
than the tip of the red giant branch magnitude at the distance of M81, 
and with colours redder than the reddest colours predicted for red 
supergiant stars. The criteria used to identify young star candidates 
were defined by inspecting the foreground-corrected CMD and the predicted 
locations of young stellar populations at the distance of M81. 
The selection boxes of young stars are shown in Fig.\,\ref{cmd_fg_corrected}. 
Objects within the blue selection box are defined as main sequence star 
candidates, and those in the red selection box as red supergiant star 
candidates. The blue selection box is affected by a modest contamination 
from foreground Galactic stars and background galaxies, which tend to have 
red colours, than the red selection box. Despite the selection criteria 
imposed to select samples of young stars free of background contamination, 
the red side of the CMD is expected to suffer still from contamination 
(see \citet{davidge08} for more details). We have therefore adopted red 
supergiant star candidates to be only secondary tracers of the debris field. 

A potential source of errors when identifying bright stars in distant objects is 
blending. As discussed in \citet{davidge08}, who had also used the 
MegaPrime/MegaCam imager to investigate the bright stellar content of the 
M81 group, the ages of the young stellar components estimated for 
Holmberg IX, BK 3N, and Garland from MegaCam CMDs are consistent with 
those obtained from CMDs obtained with the Hubble Space Telescope 
\citep{makarova02}, suggesting that  blending is not severe, if any. 
The distribution of the young stellar clumps detected in the intergalactic 
environment along H{\sc i} tidal arms rather than been randomly distributed 
within the field supports this conclusion.
By comparing MegaCam images with those obtained with the Advanced Camera 
for Surveys on board the Hubble Space Telescope, \citet{davidge08} had shown 
that the bright blue sources in the MegaCam images are consistent with being 
single main sequence stars. In addition, for the blue objects detected in our images 
to be the results of blending, the blend stars have to be bright in $g$-band yet 
faint in $i$-band, which is inconsistent with the fact that the bulk of stars in these 
remote regions of galaxies are intrinsically faint and red.

\begin{figure*}
\includegraphics[clip=,width=0.85\textwidth]{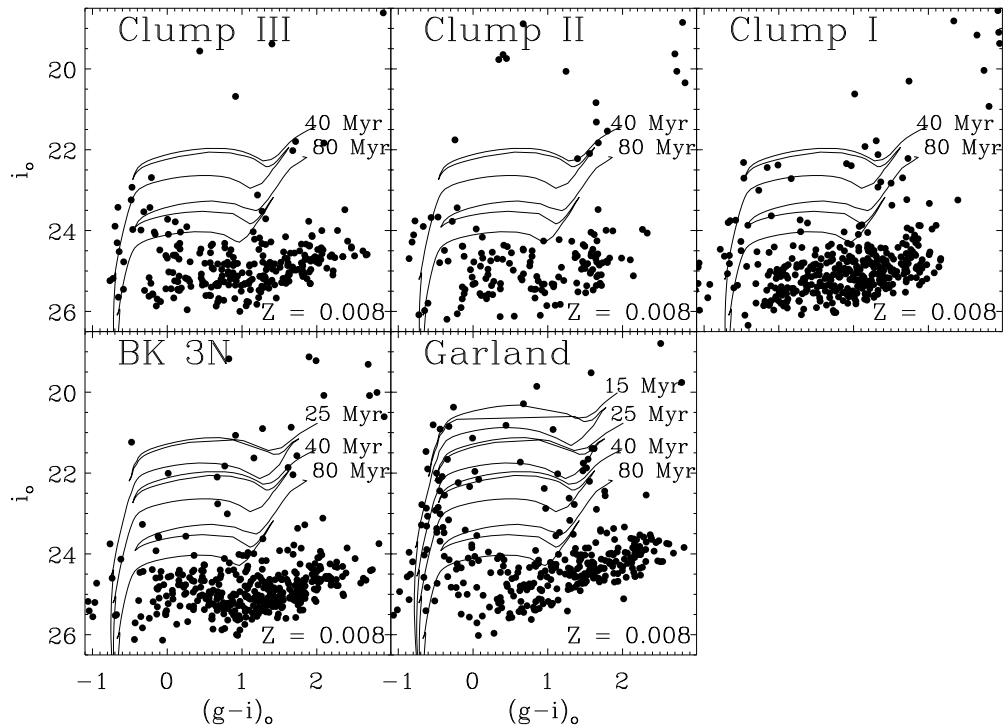}
\caption{The CMDs of the stellar clumps identified along the southern H{\sc i} 
tidal arm between M81 and NGC~3077, along with BK 3N and Garland. 
Superimposed on each CMD are the Padova isochrones with a metallicity 
Z=0.008 and the indicated ages. }
\label{cmd_td}
\end{figure*}

\begin{figure}
\includegraphics[clip=,width=0.5\textwidth]{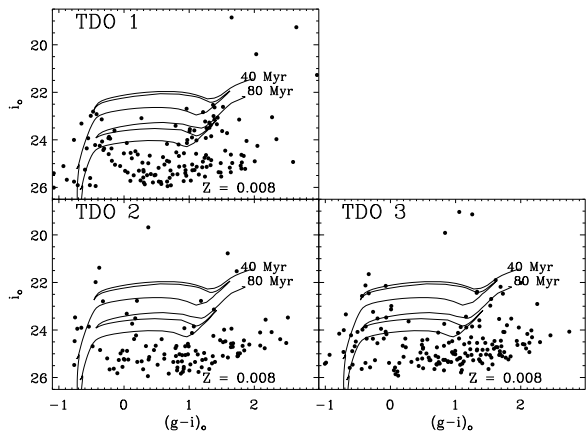}
\caption{The CMDs of the young stellar grouping identified by \citet{davidge08}
in the M81 tidal debris. Superimposed on each CMD are the Padova isochrones 
with a metallicity Z=0.008 and the indicated ages. }
\label{cmd_tdo_davidge08}
\end{figure}

Figure \ref{young_stars_map} shows the spatial distribution of main sequence 
star candidates (left panel), and both main sequence and red supergiant 
candidates (right panel) over the surveyed area. The contours and the 
grayscale image show the H{\sc i} radio map from \citet{yun94}. The crowding 
is such that individual stars are not well resolved in the central dense 
region of M81 and NGC~3077, and their photometry is affected by large errors. 
As expected, the bulk of young stellar populations are distributed over the 
star-forming disk of M81. Beyond the young disk of M81, a number of stellar 
concentrations distributed over a much larger scale are visible, of which a 
few have been already reported in the literature, i.e., Holmberg IX, BK 3N, 
Garland, M81-West, Arp's Loop (A0952+69), and the tidal debris objects 
recently reported by \citet{davidge08}. In addition to the structure combining 
M81-West and the first object in the list of \citet{davidge08}, tracing the 
continuation of the H{\sc i} spiral structure of M81, a second arm of young 
stars (M81-West2) tracing the continuation of the second H{\sc i} spiral 
structure is visible. Similar structures of young stars tracing the 
continuation of the H{\sc i} spiral structures (M81-SE) are observed to the 
south along the major axis. Wide field GALEX images of M81 \citep{gildepaz07} 
show that these structures of young star candidates are sites of Ultra-Violet 
emission.
More strikingly, a number of previously unknown overdensities of young star 
candidates can be seen along the tidal bridge of gas stretching from M81 to 
its neighbour NGC~3077. Interestingly, BK3N and Garland appear to lie at 
the two extremities of this chain of concentrations of young stars, close to M81 
and NGC~3077 respectively. The southern H{\sc i} arm has been resolved 
into a series of clouds with sizes of approximatively 4 kpc, which are embedded 
in a more tenuous H{\sc i} distribution \citep{vdh79}. The new young stellar 
systems to the south of M81 appear to trace the overdense clouds along this 
H{\sc i} arm, with no clear bridge of young stars connecting these systems.

The foreground-corrected CMDs of three control fields sampling similar 
areas on the sky as the new stellar clumps, and that are located away from 
the M81-NGC~3077 and M81-M82 debris fields, are shown in 
Fig.\,\ref{cmd_control_field}. The location of the control fields on the sky 
are shown in Fig.\,\ref{footprint}. The selection boxes of both main sequence 
and red supergiant star candidates are overplotted on the CMDs of the control 
fields. The stellar contents of the control fields are all similarly dominated by 
objects fainter than $g_{\circ}\ga 25$, i.e., red giant branch stars and 
probable background compact galaxies, with a noticeable absence of both 
stars with $(g-i)_{\circ} \la 0$, i.e., young main sequence star candidates, 
and stars with properties similar to those expected for red supergiant stars 
at the distance of M81, i.e., $(g-i)_{\circ} \sim 1.5$ and $i_{\circ} \sim 22$. 

Fig.\,\ref{cmd_sp} shows the CMDs of the stellar populations that trace the 
continuations of the H{\sc i} both to the north and to the south of the galaxy. 
Superimposed on the CMD of each stellar system as indicated in each panel 
are the Padova isochrones with a metallicity Z=0.008 and the indicated ages. 
The locations of both main sequence stars and red supergiant stars in the 
CMDs are best matched by Z=0.008 and Z=0.004 isochrones. 
The plumes of bright main sequence stars are also reasonably matched 
by isochrones of both higher and lower metallicities, however the predicted 
locations of red supergiant stars are either bluer or redder than observed. 

Fig.\,\ref{cmd_td} shows the CMDs of the stellar clumps identified along the 
southern H{\sc i} tidal tail, and the CMDs of BK 3N and Garland, both located 
at the two extremities of the same gaseous arm. Overplotted are the Padova 
isochrones with a metallicity Z=0.008 and the indicated ages. All the stellar 
systems were taken to be situated at the same distance as M81. 
Compared to the control fields, the CMDs of the stellar clumps along the 
southern H{\sc i} tidal tail show clear over-densities of objects with luminosities 
and colours expected for main sequence and red supergiant star candidates.
Similar to the stellar populations tracing the continuations of the H{\sc i} spiral 
structures, the CMDs of stellar systems tracing the southern H{\sc i} tail are best 
matched by isochrones with metallicities ${\rm Z \sim 0.008}$. A similar conclusion 
has been derived for the young stellar systems in the tidal bridge connecting M81 
and M82 \citep{demello08a}, and is consistent with the gas-phase metallicity of 
an H{\sc ii} region in Holmberg IX as measured by \citet{makarova02}. 

The CMDs show that the three newly reported stellar systems, similar to BK 3N 
and Garland, contain stars with properties consistent with a range of ages, 
i.e., they are not simple stellar populations. \citet{davidge08} derived a similar 
conclusion for three other young stellar systems of comparable extents and 
stellar densities located respectively close to the area identified as M81-West 
by \citet{sun05} and Holmberg IX. The ages of the youngest stars in 
BK 3N (Garland) are consistent with the absence (the presence) of associated 
H$\alpha$ emission \citep{karachentsev07}. Stars with photometry that are 
best matched with isochrones of ages older than a few hundred Myr, or even 
a few billion years, are present in abundance. We cannot however be sure 
that these stars are genuine members of the stellar systems, whether they 
belong to the stellar halo of M81, or whether they were ejected into the 
intergalactic medium during the interaction.

The CMDs of the stellar structures tracing the continuation of the H{\sc i} spiral 
structures of M81 (M81-West2, M81-SE) and Holmberg IX show the presence 
of stars younger than these in the stellar over-densities distributed along the 
southern H{\sc i} arm. The CMDs of the new stellar clumps show the presence 
of stars with photometric properties consistent with their youngest stars formed 
around 40\,Myr ago. BK 3N, located at the western end of the H{\sc i} tidal arm, 
appears to be dominated by stars of similar ages to the other three clumps 
along the tidal arm. The Garland structure, at the easter end of the H{\sc i} tidal 
arm, however contains stars as young as those observed in Holmberg IX and 
the spiral arms. This suggests that the star formation activity was likely truncated 
earlier in the stellar systems within the tidal field that are away from large 
galaxies than in those in their close vicinity.

The $i_{\circ}$ vs. $(g-i)_{\circ}$ CMDs of the recently reported stellar groupings 
in the debris field of M81 by \citet{davidge08} are shown in 
Fig.\,\ref{cmd_tdo_davidge08}. The locations of these stellar groupings on the 
sky are indicated in Fig.\,\ref{footprint}. Superimposed on the CMD of each stellar 
system as indicated in each panel are the Padova isochrones with a metallicity 
Z=0.008 and the indicated ages. Similar to other young stellar systems within the 
debris field of M81, the CMDs of those stellar systems are best matched by 
isochrones with metallicities ${\rm Z \sim 0.008}$. The photometric properties of 
turn-off stars in those groupings are consistent with their youngest stars formed 
$\sim 40$\,Myr ago, consistent with the age estimates of \citet{davidge08}. 
These stellar systems appear to be dominated by stellar populations similar to 
those of the stellar clumps distributed along the southern H{\sc i} arm.

\section{Discussion}
\label{discussion}

Numerous multi-wavelength data sets have unambiguously identified localised 
regions with signs of current and/or recent star formation activity distributed along 
H{\sc i} tidal tails in interacting systems \citep[e.g.][]{weilbacher03,hibbard05,demello08b},
with a number of these regions suspected to be bounded, the so-called tidal dwarf 
galaxies \citep[e.g.][]{duc98,braine01,hancock09}. The star formation activity along 
the tidal tails appears to be distributed with a similar morphology to H{\sc i} 
\citep[e.g.][]{hibbard05,neff05,hancock07}, in agreement with our finding of the new 
stellar clumps reported here tracing the dense regions along the southern H{\sc i} 
tidal tail.
\citet{hibbard05} have found that UV colours of localised regions of star formation 
along the tidal tails of the archetypal merging system NGC 4038/39 are consistent 
with continuing star formation. \citet{neff05} have argued however, for the case of 
NGC 7769/71, NGC 5713/19, and the NGC 520 system, that most of young stars in 
the tails have most likely formed in single bursts. The CMDs of stellar clumps in the 
debris field of M81, e.g. Arp's loop region, Holmberg IX, contain stars with photometric 
properties consistent with a wide range of ages,  i.e., from $\sim10\,$Myr to 
$\sim1\,$Gyr \citep[e.g.][]{demello08b,sabbi08}, suggesting extended star formation 
histories. Deep imaging data show however a lack of any concentration of old stars 
associated with the blue stars, suggesting that the ``old'' stellar component seen in 
the CMDs of those stellar clumps were formed in the stellar disks of M81 and ejected 
into the intergalactic medium during tidal passages, whereas the young stars have 
formed in the tidal debris \citep{demello08b, weisz08}. The spatial distribution of 
red giant branch stars over the region connecting M81 and NGC~3077 does not 
show any noticeable concentrations of old stars associated to the young stellar 
clumps identified along the southern H{\sc i} tidal tail. As for, e.g. Holmborg IX and
Arp's loop region, this suggests that the old stars seen in the CMDs of the newly 
reported stellar clumps should have come from one of the interacting systems while, 
since the stellar clumps along the southern tidal tail are considerably younger than 
its dynamical age, e.g. $\sim 250\,$Myr \citep{yun99}, young stars formed on site. 

Fitting single stellar population synthesis models to UV/optical colours of UV-bright 
stellar substructures within the tidal tails of four ongoing galaxy mergers, 
\citet{neff05} found that the star formation appears to be older near the parent 
galaxies and younger at increasing distances. They have suggested that this could 
be because the star formation occurs progressively along the tails, or because the 
star formation has been inhibited near the galaxy/tail interface. 
\citet{hibbard05} had reported negative UV and optical colour gradients along 
the tidal tails of the ``Antennae'' system, indicative of negative age gradients when 
moving outward along the tails  \citep[see also][]{hibbard01}. Note that the observed 
colour gradients could be accounted for alternatively by the presence of composite 
stellar populations. The star formation within the tidal tails in M81 debris field 
appears to be different. The CMDs of the bulk of stellar clumps in the debris field 
of M81 indicate that they have ceased forming stars at similar epochs in the past, 
i.e., $\sim40$\,Myr ago. This suggests that the star formation throughout M81 tidal 
debris field could have been triggered by common events 
\citep[see][for a similar conclusion]{davidge08}, and that the physical conditions 
within the dense regions along the southern H{\sc i} tidal tail are comparable. 
The systems within the debris field with younger stars, i.e., Holmberg IX and Garland, 
are both in the close vicinity of M81 and NGC~3077 respectively, in contrast with the
findings of \citet{neff05}. The diversity of these star formation histories could be most 
likely related to different gas contents and conditions within the tidal tails of those 
interacting systems.

The exact nature of the previously known young stellar systems within the tidal 
field of the M81 group, i.e.,  Holmberg IX, Gerland, and BK 3N, is not entirely 
clear yet. Detailed modelling of the dynamics of the M81 group suggests that 
the three largest galaxies in the system had an interaction $\sim 250$ Myr ago 
for M81 and NGC~3077, and $\sim 200$ Myr ago for M82 and M81 \citep{yun99}. 
The modelling of optical CMDs of Holmberg IX, BK 3N, and Arp's loop, of 
comparable depth to ones presented here, suggested that these galaxies have 
experienced star formation between about 20 and 200 Myr ago \citep{makarova02}. 
It has been argued then that these galaxies are tidal dwarf condidates that 
formed from dust and gas that was blown away from M81 and/or other galaxies in 
the group \citep{boyce01,makarova02}. BK3N may be alternatively a pre-existing 
dwarf irregular galaxy undergoing an interaction with M81 \citep{boyce01}. 
The  old stellar population associated spatially to Holmberg IX is suspected to 
belong quite likely to the outer regions of M81, suggesting that this stellar system 
is of a tidal origin \citep{sabbi08}.
 
The ages and metallicities of the isochrones that best reproduce the CMDs of
the stellar systems along the southern H{\sc i} arm are similar to these needed
to account for the properties of the stellar contents of previously known tidal 
dwarf candidates in the tidal debris field. This suggests that they both could 
share similar star formation histories and might be associated to similar events. 
These isochrone metallicities are significantly higher than the typical 
metallicities of dwarf galaxies of comparable luminosities \citep[e.g.][]{rm95}. 
This suggests that these stellar systems have been assembled from pre-enriched 
material. This is consistent with the conclusions of \citet{boone05} who found 
that the abundances and physical conditions of the molecular complex situated 
near the line of sight toward Holmberg IX are similar to those found in the disks 
of spiral galaxies. The distribution of the stellar clumps along an H{\sc i} tail, 
tracing the densest clouds within the gaseous arm, indicates that these systems 
may have been assembled out of gas pulled from one of the large interacting 
galaxies in the group. A primary criterion for the determination of the nature of 
these objects is to measure their mass-to-light ratio, which tend to be low for 
tidal dwarf galaxies due to the absence of dark matter \citep[e.g.][]{BH92, duc00}. 
Unfortunately this cannot be measured from the dataset in hand. Without this 
measurement, we can only conclude that the newly reported stellar clumps are 
likely (among the nearest) tidal dwarf galaxies.

\begin{table}
\caption{Coordinates of the new young stellar clumps along the
southern H{\sc i} arm. } 
\label{gal_prop_obs1}
\begin{tabular}{lll}
\hline
ID         &  R.A. (J2000)  & Dec. (J2000) \\
\hline
Clump I    & 09:57:21.2  &  68:42:55  \\
Clump II   & 09:59:40.4  &  68:39:19\\
Clump III  & 10:00:40.4  &  68:39:37  \\
\end{tabular}
\end{table}

\section*{Acknowledgments} 
We would like to warmly thank Mike Irwin for (various) helpful discussions.

\bsp

\label{lastpage}

\end{document}